# Free-Oscillations Coincident with Earthquakes

Randall D. Peters
Physics Department, Mercer University
Macon, Georgia

## Abstract

Two de-modulation algorithms are used to observe free-oscillations of the earth that are coincident with earthquakes.

**Background**

After the Great Chilean earthquake of 1960, it has become commonplace to study the eigenmode oscillations of the Earth after large earthquakes. Consider the physics relevant to any understanding of the mechanisms responsible for their excitation. It is illogical to believe that the genesis of these oscillations could somehow lag the occurrence of the earthquake by a significant amount of time. Moreover, some mode(s) could very possibly serve as a trigger mechanism for the earthquake to begin with.

So an obvious question begs for an answer. Why are these modes not immediately obvious in the power spectra of records containing an earthquake? The answer must involve properties of the multiplicity of complex (including nonlinear) motion-types that accompany the rupture. These are necessarily distributed through a very broad range of frequencies, as expected from even linear homegeneous elasticity theory. The large bandwidth requirement for their complete study results from the short-duration of the primary rupture, which approximates a delta function.

One might assume that the absence of any obvious eigenmode lines in a spectrum, obtained from an adequately sensitive seismograph, derives from some deficiency of the Fourier transform algorithm used to generate the spectrum. The author of the present article believes otherwise and his conclusions derive from studies of a different system that is now described.

**Seismocardiography**

As with a seismograph to study the earth, an accelerometer, such as a geophone placed on the chest of a subject, can provide valuable information concerning the heart of that subject. When the geophone records show obvious time-domain periodicity for both heart-beat and respiration, associated spectral lines would seem by most to be logically expected. In fact, they are oftentimes missing from a conventional spectrum, and the same result is found also to be true, even of spectra generated from records of electrocardiographic type.

**Remedy**

Some physiologists who were puzzled by the complexities of their spectral lines have found a method with which to `extract' information otherwise missing from their frequency-domain records. They began to work with the so-called `Teager-Kaiser Energy (TKE)' operator. Although the mechanisms responsible for various observed phenomena have not been explained, it is certainly the case that they are being influenced by modulation processes. Some investigators even refer to the TKE operator as an AM-FM demodulator [1].

**De-modulation**

Because of his cardiac studies, the present author decided to look also for modulation-effects in earthquake records. The first de-modulation technique used for this work was of the type employed by AM-radio receivers. It is a rectification scheme in which all negative-going parts of the zero-mean (raw) signal are simply set to zero before the FFT is calculated. The second technique is a modification of the TKE operator, which the author has labeled TKP detector (or tracker), where `P' stands for `peak'. Both of these de-modulators are discussed in detail in Ref. [2].

**Results**

All of the time traces shown in the figures that follow are from the output of a VolksMeter [3] using a (sub) sampling rate of 1 Hz. Most of the records, which show an earthquake, are of 24-h duration; so each time trace contains 86,400 points. The last case, for which `eq' is not included in the time trace label, is a quiet time record for comparison purposes. With the exception of the two New Zealand earthquake records, the instrument responsible for all of the data is the two channel VolksMeter situated on the seismic pier in the Physics Department of Mercer University in Macon, Georgia. The VolksMeter that recorded the New Zealand event is located in Coonabarabran, Australia at the Edward Pigot Seismic Observatory. Those show both the North/South and East/West response of the instrument. All other cases show only the East/West acceleration.

**Concerning the figures**

   Time traces:

The CDC count is a measure of pendulum angular deflection due to acceleration, with a calibration constant for the VolksMeter of $2.5 \times 10^9$ cts/rad.

   Density spectrum plots:

The first six figures show a high degree of correlation between various spectral lines of (i) the TKP detector output and (ii) the rectified signal output. These highly-correlated line regions are highlighted by the red symbols; it is concluded that they correspond to eigenmode oscillations that were hidden from view in conventionally generated spectra. The de-modulation process has made them visible, along with other lines that are assumed to be noise. Note that in the first six figures the TKP detector shows frequency characteristics similar to a low pass filter, so that components higher in frequency than about 40 mHz are not seen in these linear-ordinate graphs; though they show up clearly in the plots generated from the rectified signal.

In the last Fig. 9 a characteristic of the TKP detector is quite obvious. When dealing with low signal to noise ratios, noise increases with frequency as compared to the rectified signal. This is just opposite to the trend observed in the earlier figures.

**Conclusion**

When the SNR is not below a yet-to-be-quantified threshold, free-oscillations of the Earth can be seen in records containing an earthquake. This is accomplished by looking for spectral lines having identical frequencies in the two density spectrum plots described in the present article.

**FIGURES**

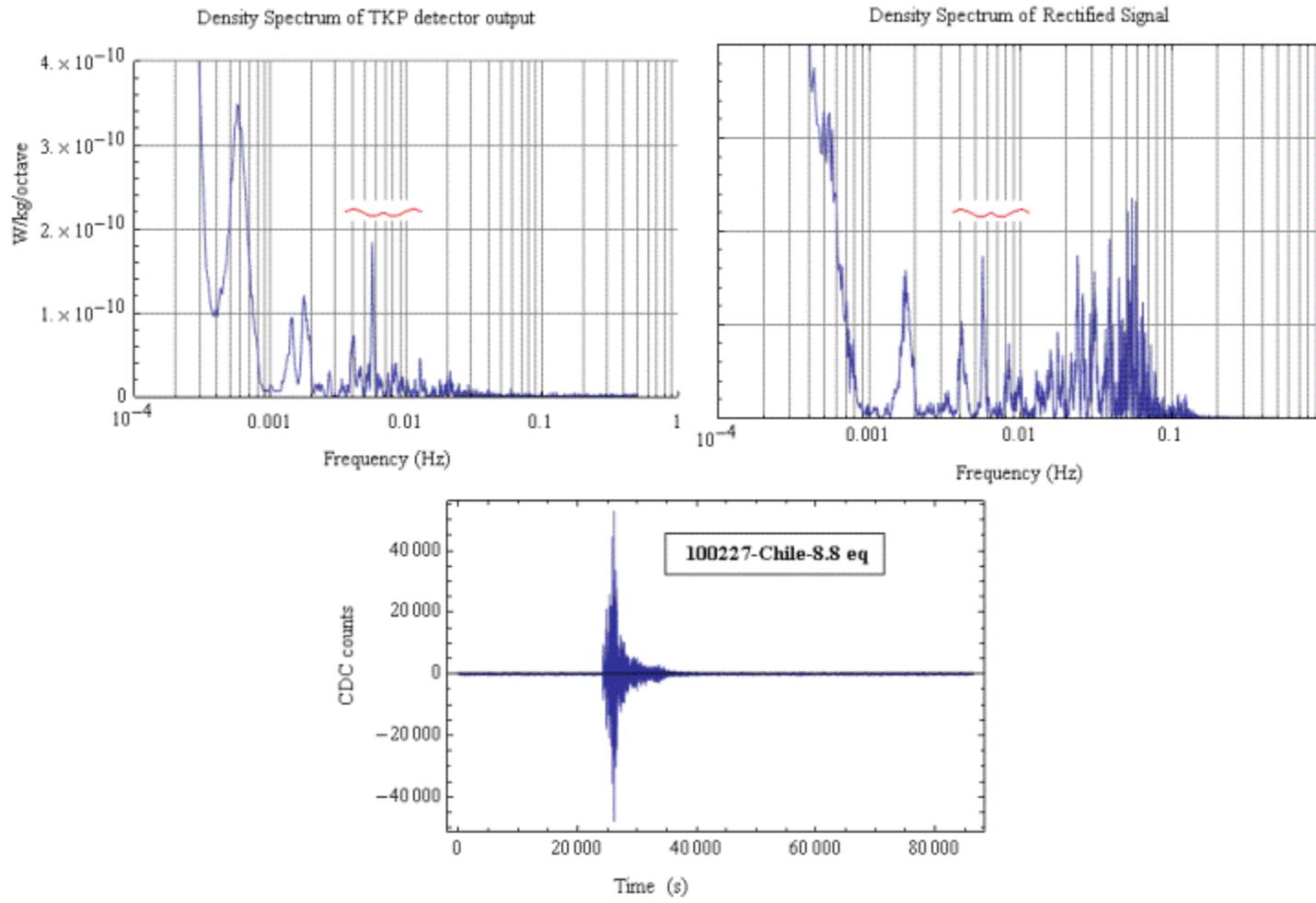

**Figure 1** Chile earthquake of 27 Feb 2010, M 8.8.

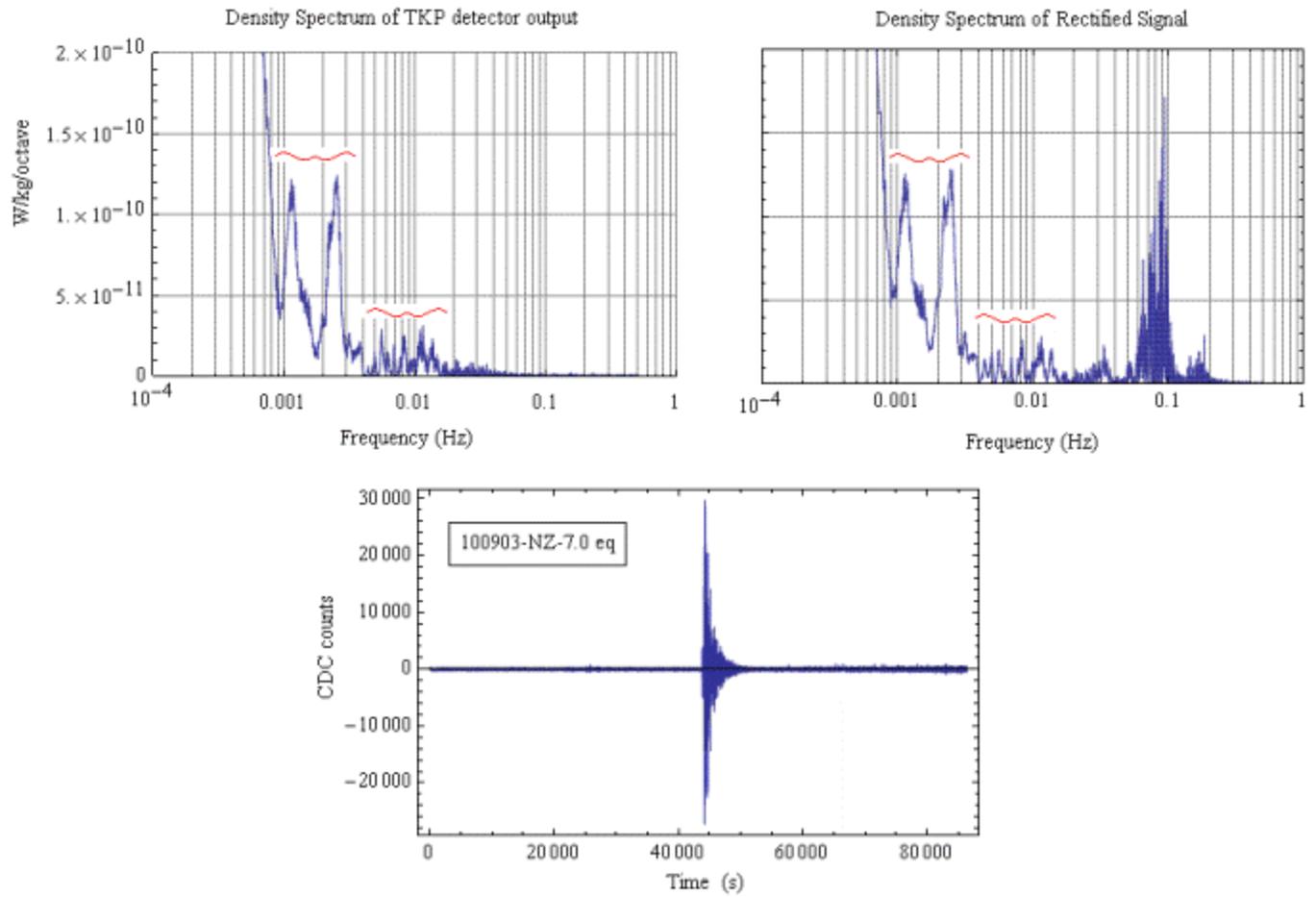

**Figure 2** New Zealand earthquake of 3 Sep 2010, M 7.0, output from the N-S channel of the VolksMeter seismometer in Coonabarabran, Australia.

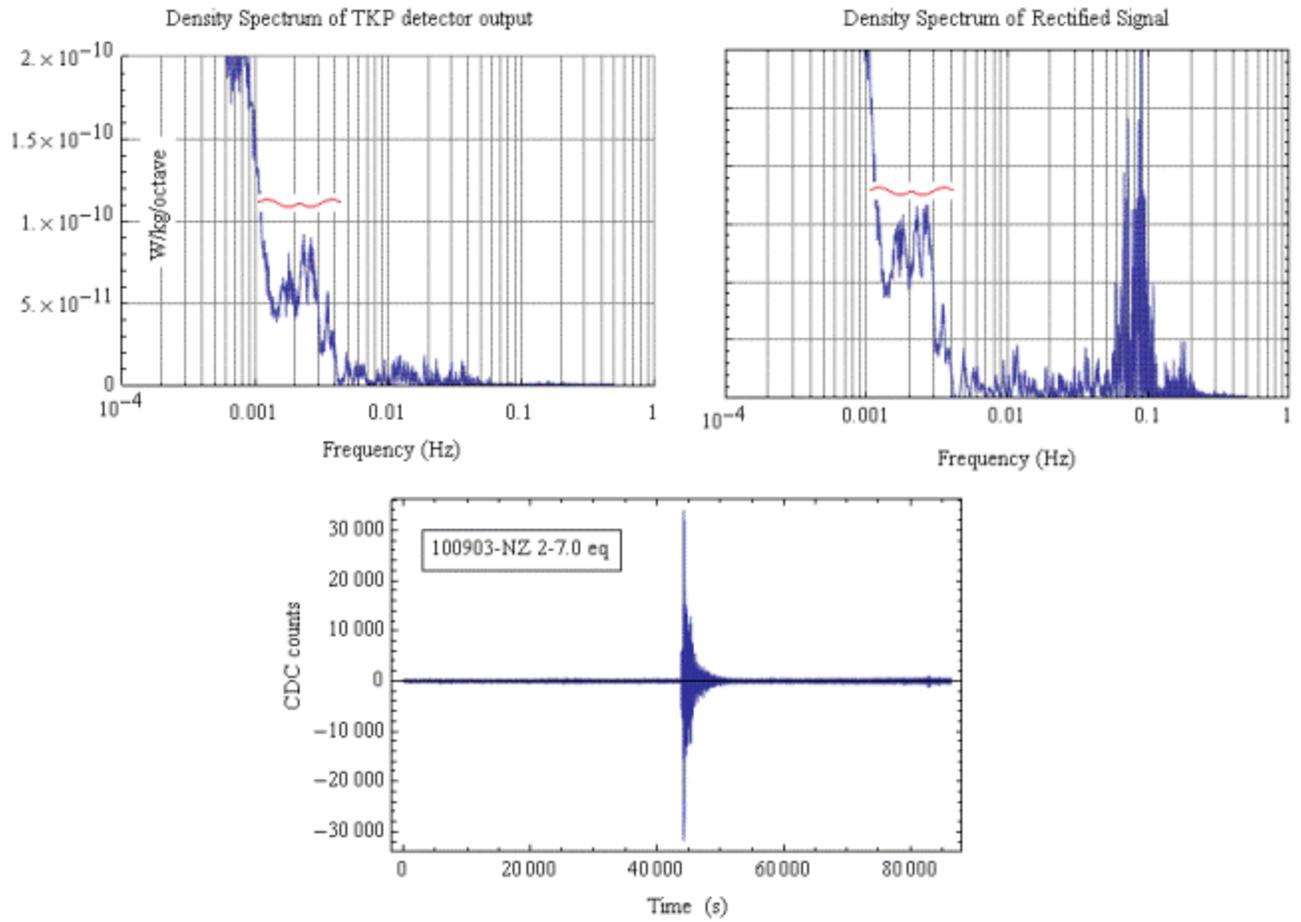

**Figure 3** Same as Fig. 2 except E-W channel.

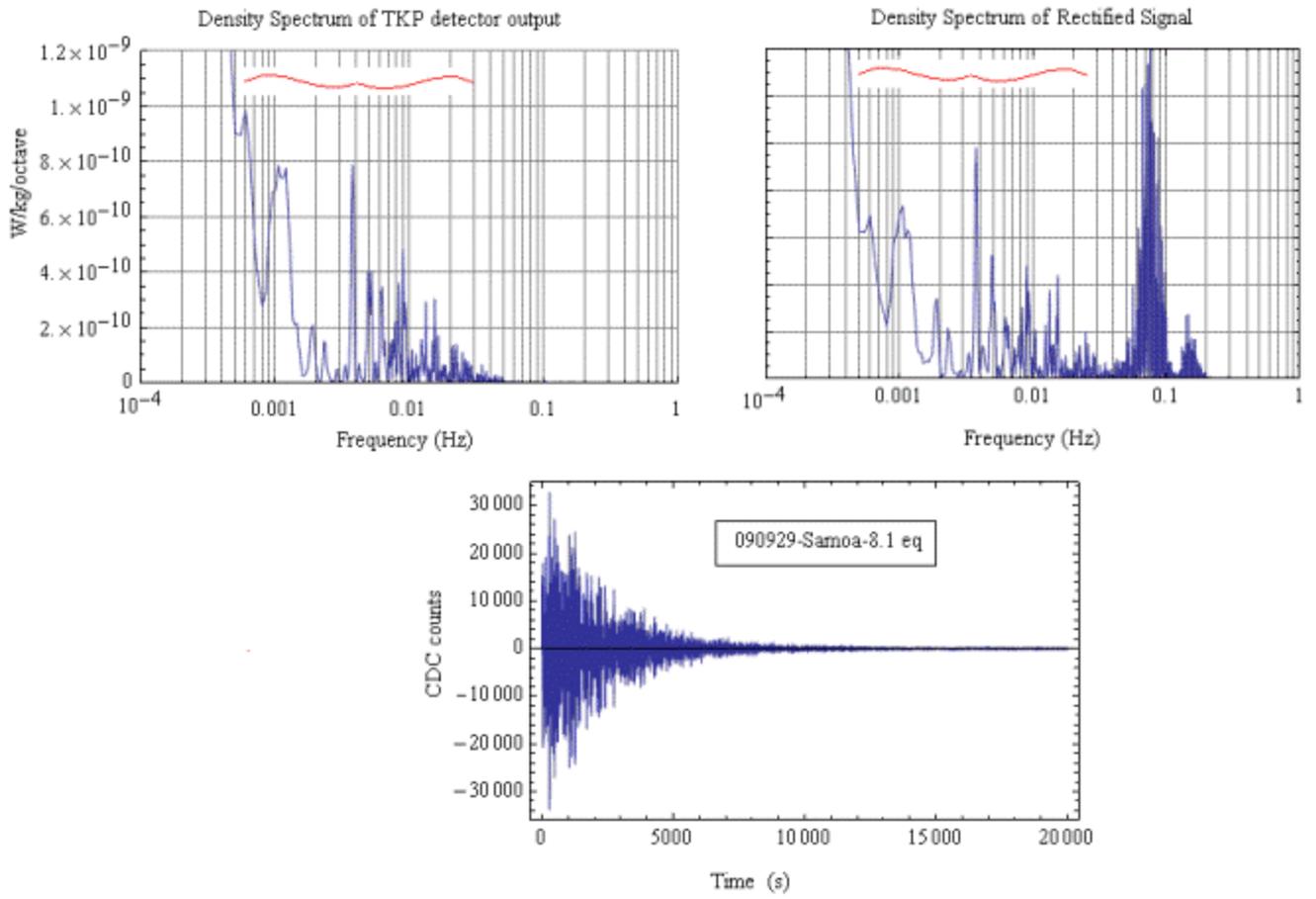

**Figure 4** Samoa earthquake of 29 Sep 2009, M 8.1.

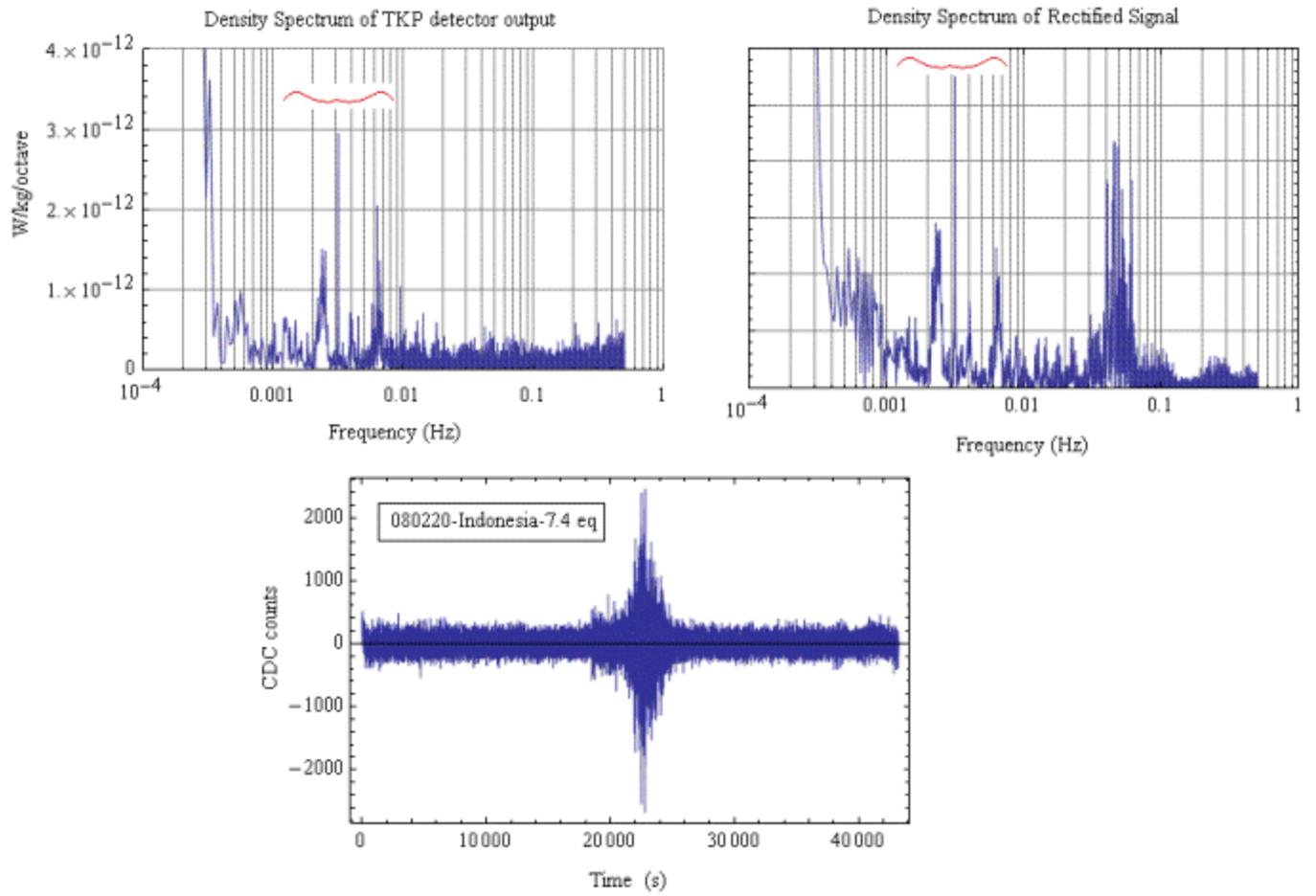

**Figure 5** Indonesia earthquake of 20 Feb 2008, M 7.4.

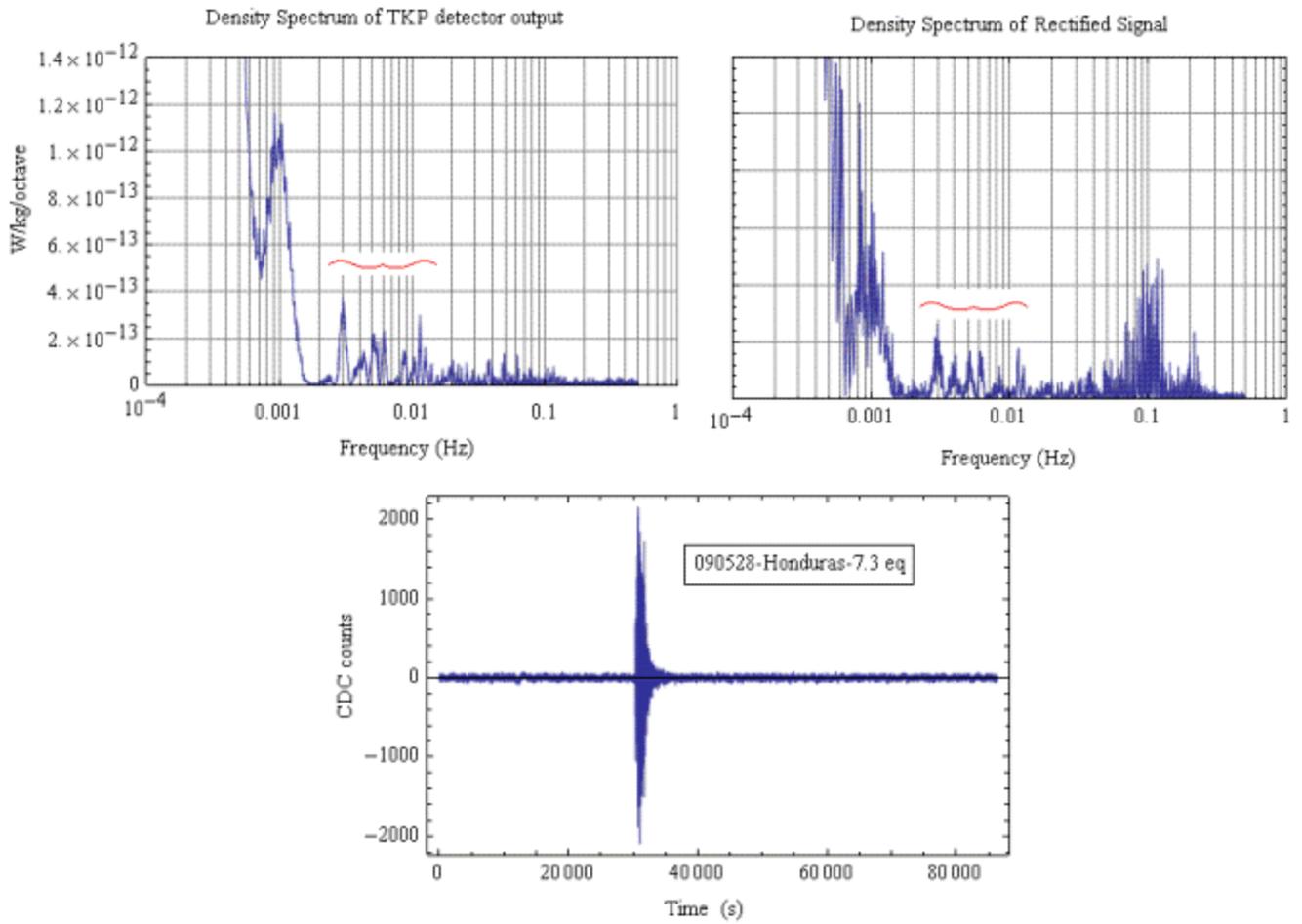

**Figure 6** Honduras earthquake of 28 May 2009, M 7.3.

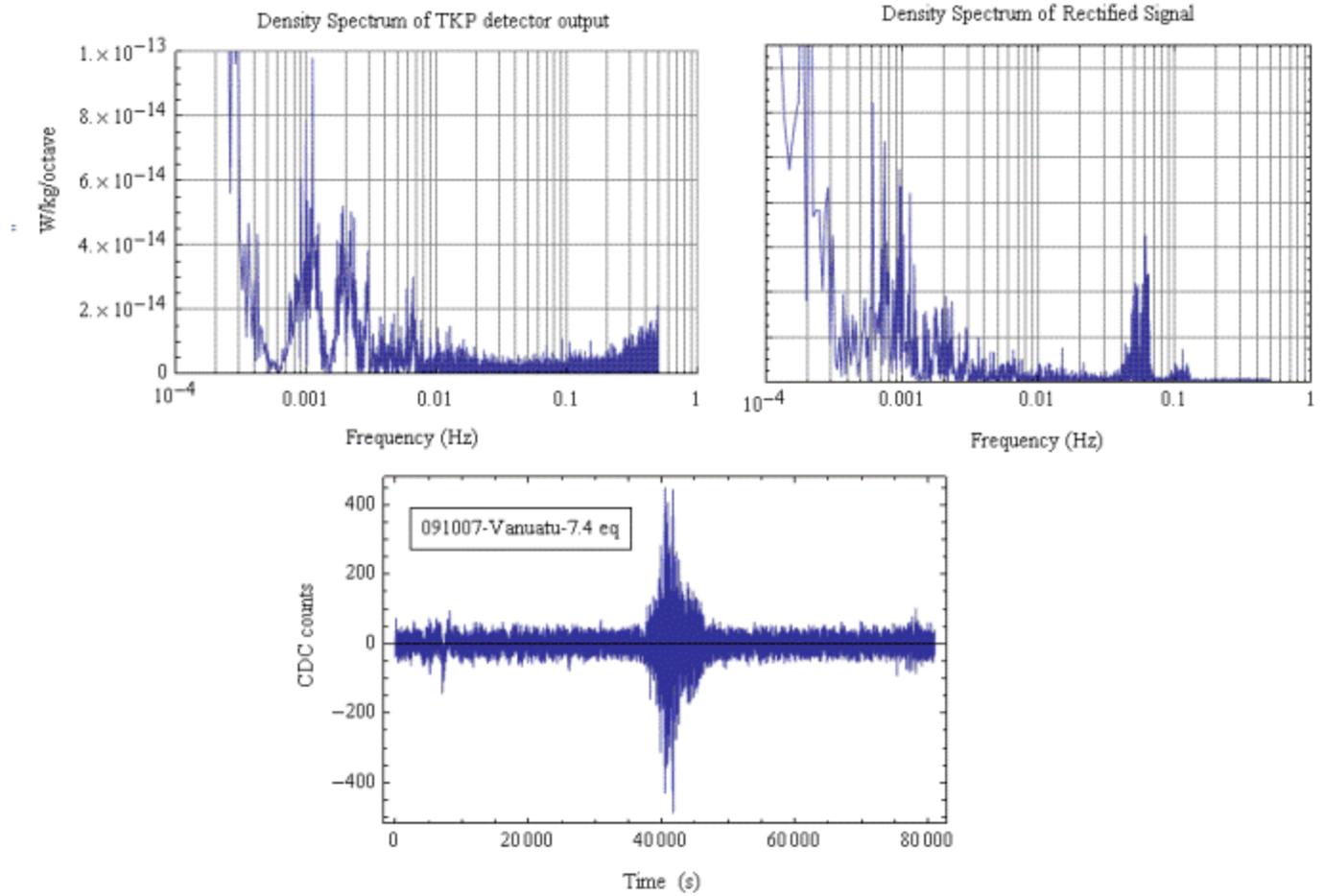

**Figure 7** Vanuatu earthquake of 7 Oct 2009, M 7.4. Correlation between the two spectra is poor. Note the peak CDC counts level at about 400.

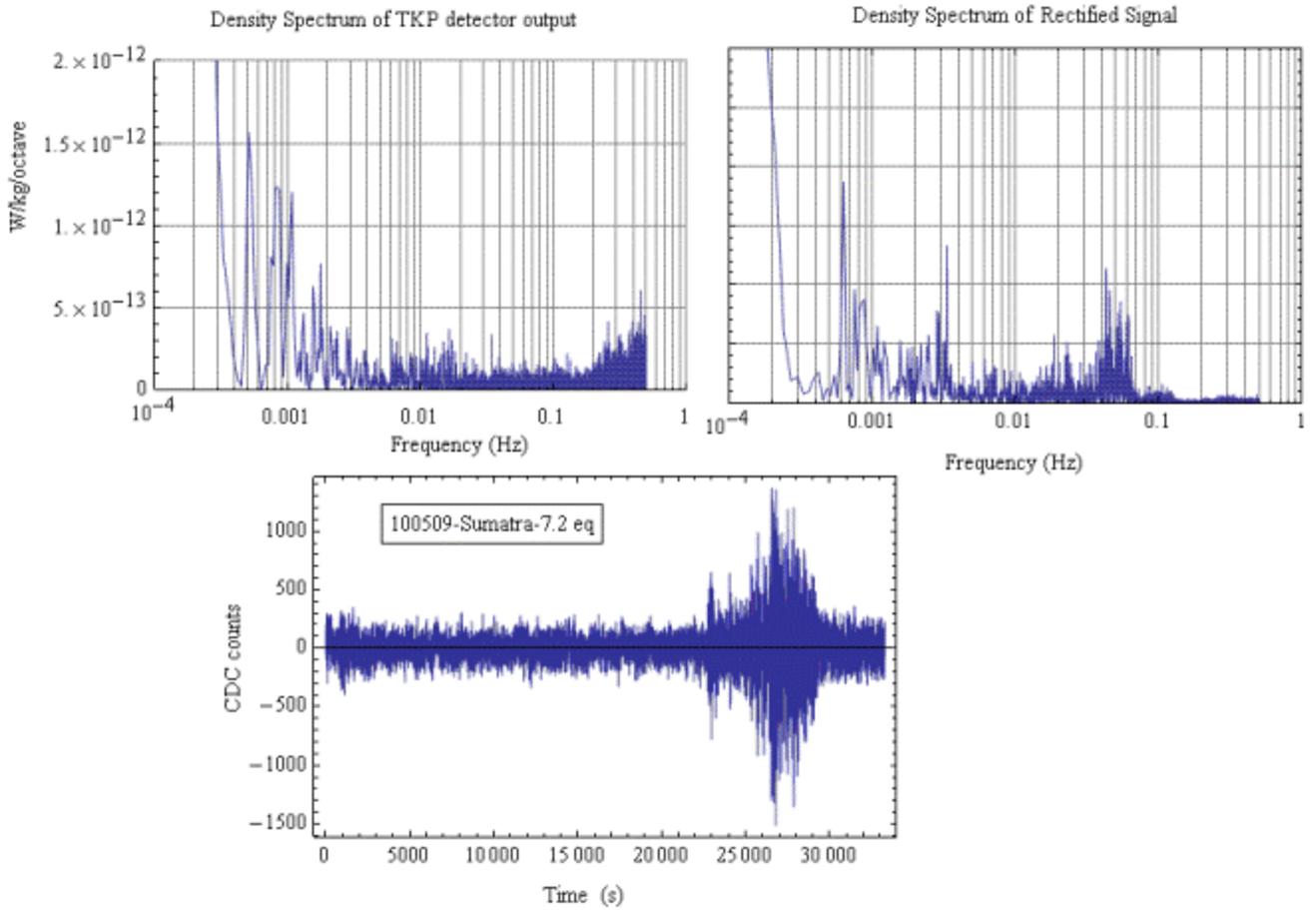

**Figure 8** Sumatra earthquake of 9 May 2010, M 7.2. Correlation between the two spectra is poor. Note the poor signal to noise ratio and that the earthquake occurs near the end of the record.

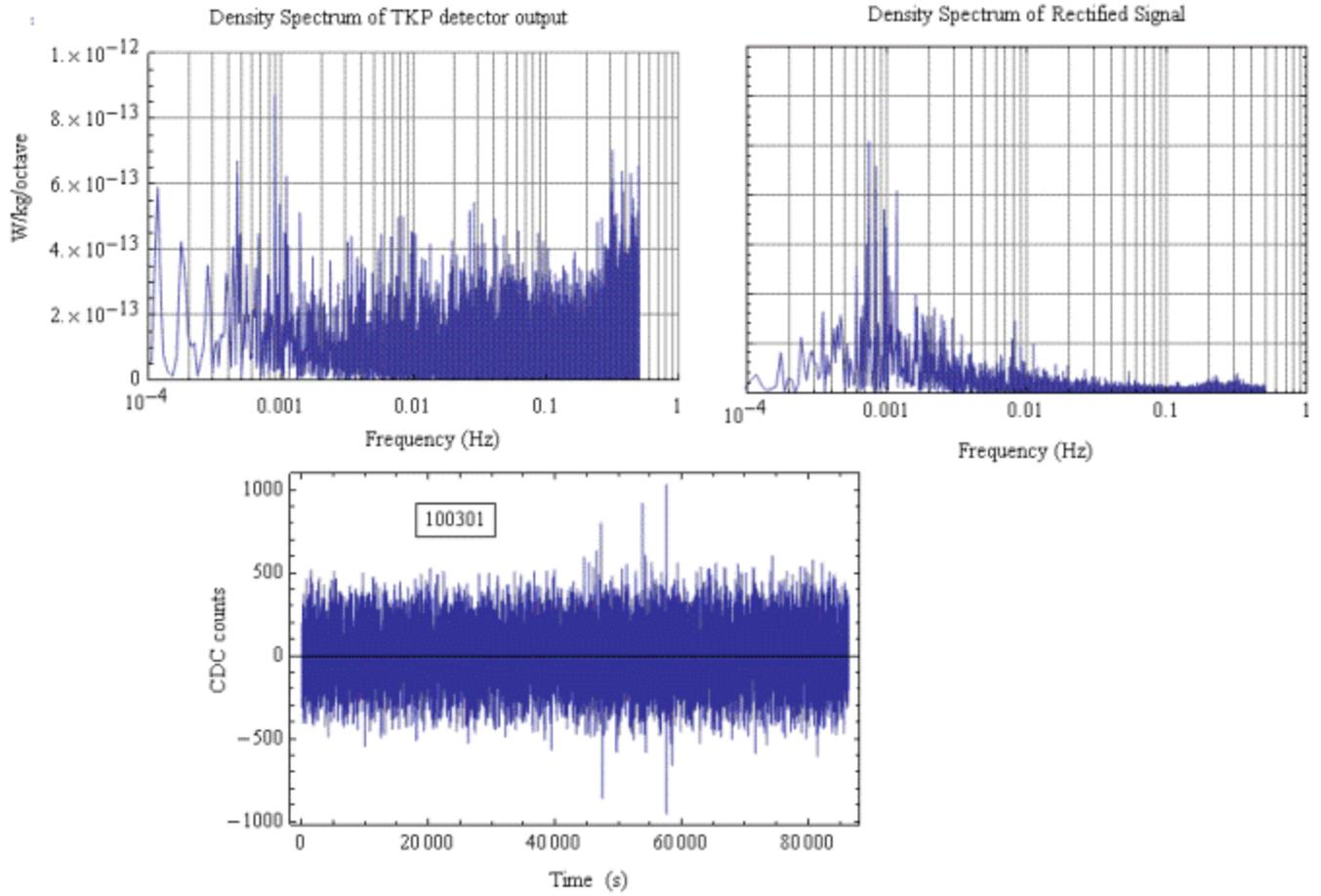

**Figure 9** A representative quiet time record. Correlation between the two spectra is poor. Note the spectral trend reversal above 10 mHz as compared to the first six figures.